\documentclass[10pt,a4paper,arial,twocolumn]{article}
\usepackage{latexsym}
\usepackage{amsmath}
\usepackage{amsfonts}
\usepackage{amsbsy}
\usepackage{amssymb}
\usepackage{epsfig}
\usepackage{mathrsfs}
\usepackage{epsfig}
\usepackage{psfrag}
\usepackage{bbm}

\numberwithin{equation}{section}

\begin{document}

\title{Operator Spin Foams: holonomy formulation and coarse graining}

\author{Benjamin Bahr\\  \emph{\small Department of Applied Mathematics and Theoretical Physics}\\ \emph{\small Wilberforce Road, Cambridge CB3 0WA, UK}}

\maketitle

\abstract{\noindent A dual holonomy version of operator spin foam models is presented, which is particularly adapted to the notion of coarse graining. We discuss how this leads to a natural way of comparing models on different discretization scales, and a notion of renormalization group flow on the partially ordered set of $2$-complexes.}


\section{Introduction}
In \cite{arXiv:1010.4787} a general class of models was presented, which are a generalization of current spin foam models (see \cite{arXiv:1004.1780} and references therein), called operator spin foams (OSF). Based on the work \cite{arXiv:0909.0939}, these models are defined on a $2$-complex $\kappa$, which can be seen as either embedded in a manifold (to depict a history of spin networks), or as abstract set of vertices, edges, faces, and boundary relations, like it would appear e.g.~as term in a GFT expansion. They are quite general, containing the Ponzano-Regge- , Barrett-Crane- and EPRL-FK model (for $\gamma<1$) as special cases.

In this article we will present a dual version of the OSF in terms of certain holonomies. Introducing a notion of coarse graining, we will also discuss how to compare models on different $\kappa$, leading to a notion of renormalization group flow.

\section{Operator spin foams}

\begin{figure}[hbt]
\begin{center}
	\psfrag{e}{$e$}
    \psfrag{f1}{$f_4$}
    \psfrag{f2}{$f_3$}
    \psfrag{f3}{$f_2$}
     \psfrag{f4}{$f_1$}
\includegraphics[scale=0.25]{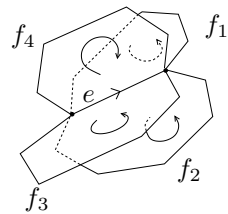}
\caption{An edge $e$ and four incident faces, with agreeing or opposite orientations.}\label{Fig:EdgeHilbertSpace}
\end{center}
\end{figure}

Let $\kappa$ be a finite, locally finite, oriented $2$-complex, and $G$ a compact Lie group. An operator spin foam consists of the following data: an assignment of irreducible, finite-dimensional, unitary representations $\rho_f$ of $G$ to faces $f$ of $\kappa$. For each such assignment, denote the Hilbert space
\begin{eqnarray}\label{Gl:EdgeHilbertSpace}
\mathcal{H}_e\;=\;V_{\rho_{f_1}}\otimes \cdots V_{\rho_{f_n}}\otimes V_{\rho_{f_{n+1}}}^*\otimes\cdots V_{\rho_{f_m}}^*
\end{eqnarray}
\noindent for each edge $e$ of $\kappa$, where $f_1,\ldots,f_n$ are the faces incident to $e$ with agreeing orientation, and $f_{n+1},\ldots,f_m$ the ones with opposite orientation (see figure \ref{Fig:EdgeHilbertSpace}). For each edge, assign an operator
\begin{eqnarray}
P_e\;:\;\mathcal{H}_e\;\longrightarrow\;\mathcal{H}_e
\end{eqnarray}
\noindent Then, at each vertex $v$ of $\kappa$, due to the index structure of the operators $P_e$, there is a canonical way of contracting the indices of all the $P_e$ in order to obtain a number, which is denoted by $\text{tr}_v$. More generally, if $\kappa$ has a non-empty boundary, which in \cite{arXiv:0909.0939} was defined as the graph $\gamma$ consisting of all edges $e$ in $\kappa$ that are bounded by only one face, then the $P_e$ are only assigned to edges \emph{not} in $\gamma$. The index structure is such that contracting over all vertices \emph{not} in $\gamma$ leads to a linear form on $\mathcal{H}_\gamma$, the space of $G$-spin networks on $\gamma$ (with representations assigned to each edge equal to the $\rho_f$ of the unique face $f$ touched by $e$).
\begin{eqnarray}
Z(\kappa,\rho_f,P_e)=\prod_v\text{tr}_v\left(\otimes_{e\supset v}P_e\right)\;:\;\mathcal{H}_\gamma\to\mathbb{C}
\end{eqnarray}
\noindent When the graph $\gamma=\gamma_i\sqcup\gamma_f^*$ is disconnected (where $\gamma_f^*$ denotes the graph $\gamma_f$ with all orientations reversed), then $Z(\kappa,\rho_f,P_e):\mathcal{H}_i\to\mathcal{H}_f$ is interpreted as the transfer operator of the theory.

If $P_e$ is chosen to commute with the group action on each $\mathcal{H}_e$, then the operator $Z(\kappa,\rho_f,P_e)$ becomes gauge-invariant, and $Z(\kappa,\rho_f,P_e)$ is invariant under trivially subdividing a face $f$ by an additional edge. If the $P_e$ are chosen to be self-adjoint, then $Z(\kappa,\rho_f,P_e)$ is invariant under change of an edge-orientation. If it furthermore commutes with the isomorphisms $V_\rho\to V_\rho^*$, it is also invariant under change of a face-orientation. If $P_e$ is chosen to be an orthogonal projector, then $Z(\kappa,\rho_f,P_e)$ is invariant under changing $\kappa$ by trivially subdividing an edge by an additional vertex.

\section{Dual formulation}

In the following we introduce a dual formulation of the OSF in terms of certain holonomies.\footnote{This formulation is a generalization of the holonomy formulations given in \cite{gr-qc/0112002} for the Barrett-Crane-, and in \cite{arXiv:1010.5227} for the EPRL-FK model.} This also allows us to speak of the sum over all $\rho_f$. Denote the set of edges in $\kappa$ by $E$ and faces in $\kappa$ by $F$. Then define the set
\begin{eqnarray}
E\ltimes F\;:=\;\big\{(e,f)\in E\times F\,\big|\,e\subset\partial f\big\}
\end{eqnarray}
\noindent The dual formulation is given in terms of integral over $G^{E\ltimes F}$, i.e.~using holonomies $h_{(e,f)}$ with $e\subset\partial f$. Define, for an edge, the function
\begin{eqnarray}\nonumber
&&C_e(h_{(e,f_1)},\ldots, h_{(e,f_n)})\;:=\;\sum_{\{\rho_{f_k}\}}\left(\prod_{k=1}^{m}\dim\rho_f\right)\\\label{Gl:DefEdgeFunction}
&&\;\times
\prod_{k=1}^n\rho_k(h_{e,f_k})^{m_k}{}_{n_k}\prod_{k=n+1}^m\rho^*_k(h_{e,f_k})_{m_k}{}^{n_k}\\[5pt]\nonumber
&&\;\times (P_e)^{m_1\cdots m_n}{}_{m_{n+1}\cdots m_m,n_1\cdots n_n}{}^{n_{n+1}\cdots n_m}
\end{eqnarray}

\begin{figure}
\begin{center}
	\psfrag{f}{$f$}
    \psfrag{e1}{$e_1$}
    \psfrag{e2}{$e_2$}
    \psfrag{e3}{$e_3$}
    \psfrag{e4}{$e_4$}
    \psfrag{e5}{$e_5$}
\includegraphics[scale=0.35]{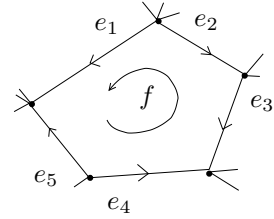}
\caption{The holonomy $g_f$ around a face $f$ is given by an ordered product of $h_{(e,f)}$, $e\subset\partial f$. In this case $g_f=h_{(e_1,f)}h_{(e_2,f)}^{-1}h_{(e_3,f)}^{-1}h_{(e_4,f)}h_{(e_5,f)}^{-1}$}\label{Fig:Curvature}
\end{center}
\end{figure}

\noindent If the norm of the operators $P_e$ does not grow too fast w.~r.~t.~the $\rho_f$, then (\ref{Gl:DefEdgeFunction}) exists as a distribution on $G^{E\ltimes F}$. For each face $f$ consider a smooth class function $S_f(g) = S_f(hgh^{-1})$, and denote its Fourier coefficients by
\begin{eqnarray}
\hat S_f(\rho)\;=\;\int_Gdg\,S_f(g)\chi_\rho(g)
\end{eqnarray}
\noindent where $\chi_\rho$ denotes the character in the representation $\rho$. Finally, for a face $f$ denote the holonomy around $f$ by (see figure \ref{Fig:Curvature})
\begin{eqnarray}\label{Gl:DefHolonomy}
g_f\;:=\;\overrightarrow{\prod_{e\supset \partial f}}h_{(e,f)}^{\pm 1}
\end{eqnarray}

\noindent Then it can be shown that
\begin{eqnarray}\nonumber
Z[\kappa]\;&:=&\;\int_{G^{E\ltimes F}}dh_{e,f}\prod_fS(g_f)\prod_eC_e(h_{(e,f_1),\ldots})\\[5pt]\label{Gl:DualFormulation}
&=&\;\sum_{\{\rho_f\}}\left(\prod_f\hat S_f(\rho_f)\right)Z(\kappa,\rho_f,P_e)
\end{eqnarray}

\noindent Note that formula (\ref{Gl:DualFormulation}), allows to write the sum over spin foam amplitudes as either a sum over representations or integral over $G^{E\ltimes F}$. The $\hat S_f$ here act as face amplitudes, and at the same time as regulators (if $S_f$ are chosen to be smooth) to make the sum finite. Note that, in the case of BF theory, when all $P_e$ are the projectors onto the gauge-invariant subspace of $\mathcal{H}_e$, one has (if all orientations of $e$ and $f_k$ agree) that
\begin{eqnarray}
C^{(BF)}_e(h_{(e,f_1)},\ldots)\;=\;\prod_{k=1}^{m-1}\delta(h_{(e,f_k)}h_{(e,f_{k+1})}^{-1})
\end{eqnarray}
\noindent so the integral over $G^{E\ltimes F}$ collapses to an integral over $G^E$. In this case, one formally also chooses $S_f(g)=\delta(g)$, so the edge amplitudes are given by $\hat S_f(\rho)=\text{dim}\,\rho$, reproducing e.g.~the known formulae for the Ponzano-Regge model.

The class of possible models which can arise by choosing different functions $S_f$ and distributions $C_e$ include not only BF theory, but also the BC- or the EPRL-FK-model (for $\gamma<1$), or the finite group spin foam models introduced in \cite{arXiv:1103.6264}.

Note that it is straightforward to include a boundary graph $\gamma\subset \kappa$, in which case the action of the operator $Z[\kappa]$ on a state $\psi$ can be either written in terms of a sum over indices $\psi^{m\cdots}{}_{n\cdots}$, or as integration over a wave function $\psi(h_1,\ldots)$. The formalism can be easily extended to include boundary edges with more than one face touching it. Details will appear in a future publication.

\section{Coarse graining}

In quantum field theories and statistical systems, coarse graining is a method to construct a theory on a larger (macroscopic) scale from a theory on a finer (microscopic) scale. The result is an effective theory which describes only the behaviour of the long range degrees of freedom within the microscopic theory. By repeated coarse graining one can construct a series of theories, labeled by their characteristic (length or energy) scale. Within e.g.~quantum field theory it is desirable to find so-called renormalizable theories. These have the same structure on all scales, differing only by a finite set of parameters, i.e.~the coupling constants, which are allowed to depend on the scale.

We now introduce a notion of coarse graining for OSF. The motivation for this is two-fold: First of all, the OSF are candidates for physical theories which are truncated to finitely many degrees of freedom by the $2$-complex $\kappa$. The setup is tailored to include background-independent theories, such as quantum gravity, so having a notion of coarse graining for these theories can address questions of renormalizability within this context. Secondly, it has been demonstrated that the discretization that lies at the foundation of Spin Foam models, breaks diffeomorphism-invariance \cite{arXiv:0905.1670}. It has also been observed in toy models that one can regain the correct diffeomorphism symmetry \emph{on the discrete level} by following the renormalization trajectory to the fixed point \cite{arXiv:1101.4775}. It is therefore important to investigate renormalization in the Spin Foam context, in order to see whether diffeomorphism symmetry can be restored this way as well.

In the following we will think of the $2$-complexes $\kappa$ as being embedded in a manifold $M$, since this induces a natural partial ordering on the set of (semianalytic) $2$-complexes. Consider two $2$-complexes $\kappa$ and $\kappa'$ such that every edge/face in $\kappa$ can be composed of edges/faces in $\kappa'$. We then write $\kappa\leq \kappa'$.

\begin{figure}
\begin{center}
	\psfrag{e1}{$e_1$}
    \psfrag{e2}{$e_2$}
    \psfrag{f}{$f$}
    \psfrag{e3}{$\!\!e'_1$}
    \psfrag{f3}{$\!\!f'_1$}
    \psfrag{e4}{$\!\!e'_2$}
    \psfrag{f4}{$\!\!f'_2$}
    \psfrag{e5}{$\!\!e'_3$}
    \psfrag{f5}{$\!\!f'_3$}
    \psfrag{e6}{$\!\!e'_4$}
    \psfrag{f6}{$\!\!f'_4$}
\includegraphics[scale=0.4]{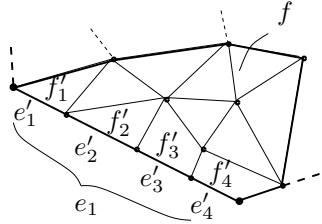}
\caption{Microscopic holonomies $h'_{(e'_k,f'_k)}$ are composed to macroscopic holonomy $h_{(e,f)}$.}\label{Fig:RelateHolonomies}
\end{center}
\end{figure}
\noindent Let $e$ be an edge in $\kappa$ composed of edges $e'_k$ in $\kappa'$. Define
\begin{eqnarray}\label{Gl:RelateHolonomies}
H_{(e,f)}\,:=\;h_{(e'_1,f'_1)}\ldots h_{(e'_n,f'_n)}
\end{eqnarray}
\noindent where the faces $f'_k$ are uniquely determined by the conditions of containing $e'_k$ and being contained in $f$ (see figure \ref{Fig:RelateHolonomies}).

Two OSF models, one defined on $\kappa$, with functions $S_f$ and $C_e$, and the other defined on $\kappa'$, with functions $S'_{f'}$ and $C'_{e'}$ are said to be \emph{cylindrically consistent} if
\begin{eqnarray}\label{Gl:CoarseGrainingCondition}
&&\mu(h_{(e,f)}):=\prod_fS_f(g_f)\prod_eC_{e}(h_{(e,f)},\ldots)\\[5pt]\nonumber
&&\;\stackrel{!}{=}\;\int dh'_{(e',f')}\prod_{f'}S'_{f'}(g_{f'})\prod_{e'} C'_{e'}(h'_{(e',f')},\ldots)\\[5pt]\nonumber
&&\quad\times\prod_{(e,f)}\delta\big(h_{(e,f)}^{-1}H_{(e,f)}\big)
\end{eqnarray}

\noindent The construction in (\ref{Gl:CoarseGrainingCondition}) is such that any function $\psi$ of the $h'_{(e',f')}$ which only depends on the combinations $H_{(e,f)}$ (\ref{Gl:RelateHolonomies}) have the same expectation values in the model defined on $\kappa$ and the one on $\kappa'$, i.e.
\begin{eqnarray}\label{Gl:CylindricalConsistency}
\int dh_{(e,f)} \mu \,\psi(h_{(e,h)})\;=\;\int dh'_{(e',f')}\mu'\,\psi(H_{(e,f)})
\end{eqnarray}

\noindent In this sense, the model on $\kappa$ is a coarse graining of the model on $\kappa'$, defining an effective theory for the macroscopic degrees of freedom. Some comments are in order:

\begin{itemize}
\item The setup can also be applied to abstract $2$-complexes, in which case one has to decide what it means that $\kappa\leq\kappa'$. In \cite{arXiv:1010.5437} a notion of refinement has been considered where $\kappa\leq\kappa'$ if each edge/face in  $\kappa$ is already contained in $\kappa'$. Note that this is a slightly weaker condition than the one considered in this article.

\item Eq.~(\ref{Gl:CoarseGrainingCondition}) can be viewed as renormalization group flow equations for OSF. If one parametrized the set of all OSF by $\{\lambda_n\}$, then (\ref{Gl:CoarseGrainingCondition}) would be equations relating $\{\lambda_n\}$ and $\{\lambda'_n\}$ on different $2$-complexes. Hence, unlike in background-dependent contexts, the $\{\lambda_n\}$ would not depend on just one length scale, but on all of $\kappa$. The RG flow would therefore not live on a single trajectory, but on the partially ordered set of all $2$-complexes.\footnote{Ideas in this direction are in fact not new, see e.g.~\cite{hep-th/0511222} for more on this issue.} Note that for these the notion of limit still exists, so it is meaningful to ask where the $\{\lambda_n\}$ are flowing to some fixed point.

\item If all the $S_f$ and $C_e$ are positive, then (\ref{Gl:CylindricalConsistency}) is in fact the condition for cylindrical consistency for the measures $\mu\, dh_{(e,f)}$ and $\mu'\,dh'_{(e',f')}$. If these conditions are satisfied on all $\kappa$, then the measures can be combined to a measure on the projective limit of all the $G^{E\ltimes F}$, which can be seen as the phase space of the continuum theory. This is in complete analogy to the construction of the space of generalized connections in LQG. In this sense renormalizability is tied to the existence of a continuum limit.


\item A priori it is not clear whether, given any OSG on $\kappa'$, (\ref{Gl:CoarseGrainingCondition}) is enough to uniquely determine the $S_f$ and $C_e$ on $\kappa$, or even that they exist at all. If not, then it might be necessary to go over to more general versions of the OSF, which also include non-localities \cite{arXiv:1111.0967}.

\end{itemize}

\section{Summary}

We have discussed a holonomy formulation for the OSF models, utilizing holonomies $h_{(e,f)}$ labeled by pairs of edges $e$ and faces $f$, with $e\subset\partial f$. These are related to the wedge holonomies. With this the OSF can be either formulated in terms of sum over representations or integration over group variables. In the latter form it gives rise to a natural coarse graining procedure, allowing it to compare models on different $2$-complexes. We have also commented on how this implies that renormalization group flow for OSF is naturally defined along a partially ordered set, rather than a single scale, and discussed how cylindrical consistency of the models is closely connected to its continuum limit.

\section*{Acknowledgements}

The author is indebted to Bianca Dittrich, Jonathan Engle, Frank Hellmann,  Wojciech Kaminski, Jurek Lewandowski and Carlo Rovelli for discussions and comments.

\end{document}